# Tunneling Enhancement of the Gas-Phase CH + CO$_2$ Reaction at Low Temperature

Dianailys Nuñez-Reyes,[1] Kevin M. Hickson,[1,*] Jean-Christophe Loison,[1] Rene F. K. Spada,[2] Rafael M. Vichietti,[3] Francisco B. C. Machado,[3] Roberto L. A. Haiduke[4]

[1] Univ. Bordeaux, ISM, CNRS UMR 5255, F-33400 Talence, France.

[2] Departamento de Física, Instituto Tecnológico de Aeronáutica, 12228-900, São José dos Campos, SP, Brazil.

[3] Departamento de Química, Instituto Tecnológico de Aeronáutica, 12228-900, São José dos Campos, SP, Brazil.

[4] Departamento de Química e Física Molecular, Instituto de Química de São Carlos, Universidade de São Paulo, 13566-590, São Carlos, SP, Brazil.

**Abstract**

The rates of numerous activated reactions between neutral species increase at low temperatures through quantum mechanical tunneling of light hydrogen atoms. Although tunneling processes involving molecules or heavy atoms are well known in the condensed phase, analogous gas-phase processes have never been demonstrated experimentally. Here, we studied the activated CH + CO$_2$ $\rightarrow$ HCO + CO reaction in a supersonic flow reactor, measuring rate constants that increase rapidly below 100 K. Mechanistically, tunneling is shown to occur by CH insertion into the C-O bond, with rate calculations accurately reproducing the experimental values. To exclude the possibility of H-atom tunneling, CD was used in additional experiments and calculations. Surprisingly, the equivalent CD + CO$_2$ reaction accelerates at low temperature as zero point energy effects remove the barrier to product formation. In conclusion, heavy-particle tunneling effects might be responsible for the observed reactivity increase at lower temperatures for the CH + CO$_2$ reaction, while the equivalent effect for the CD + CO$_2$ reaction results instead from a submerged barrier with respect to reactants.

# 1 Introduction

Typically, a chemical reaction occurs spontaneously when the reagents have enough energy for product formation to occur. Endothermic reactions and reactions which possess activation barriers over the relevant potential energy surface (PES) generally display reaction rates that increase with temperature because the higher kinetic energy leads to a greater proportion of these reagents having the minimum energy necessary for a successful collision. For these systems, an adequate description of the rate constant ($k$) can be obtained around room temperature by applying the Arrhenius' expression, $k = A\exp(-E_a/RT)$, where A is the pre-exponential parameter and $E_a$ is the activation energy (T and R refer to the temperature and ideal gas constant, respectively). Conversely, at low and very low temperatures, the rate constants for these processes are expected to become negligible since the reagents do not possess enough energy to reach the product side. While this statement is universally true for endothermic reactions, certain exothermic reactions characterized by energetic barriers have shown startling deviations from Arrhenius behaviour at low temperatures, with rate constants increasing by several orders of magnitude. This behaviour appears to be due to quantum mechanical tunneling (QMT) and has been confirmed through qualitative[1] and quantitative[2] studies of product formation. These findings are beginning to transform our view of low temperature environments such as interstellar clouds and planetary atmospheres as a wide range of activated reactions involving H/D-atom transfer that were previously neglected from models, particularly those involving OH radicals reacting with volatile organic compounds,[1,3] are potentially relevant to the overall chemistry of these regions.

Although many gas-phase reactions deviate from Arrhenius behaviour at low temperature through weakly bound complexes[4] or tunneling effects,[5,6] the reactions presenting substantial QMT effects possess specific features. Notably, a strong van der Waals complex seems to be a prerequisite for efficient QMT. Indeed, as the temperature falls, the lifetime of the molecular complex is extended, leading to an increase in the attempt frequency and an enhanced probability for tunneling to occur.[7] To date, large tunneling effects in gas-phase reactions have only been observed in systems where the mechanism involves the transfer of light atoms such as hydrogen or deuterium. Although less widespread, heavy particle tunneling effects have also been demonstrated at low temperature. In particular, carbon atom tunneling has been shown to be at the origin of several phenomena including bond shift effects in 1,3-cyclobutadiene solutions[8] and the observed ring expansion[9] of matrix isolated 1-methylcyclobutylfluorocarbene. Even molecules themselves have been shown to undergo QMT with the well-known example of formaldehyde polymerization as demonstrated by Goldanskii in the 1970's.[10,11]

Despite these earlier studies in condensed phases, there is no experimental evidence for any gas-phase reaction occurring through heavy particle tunneling yet. Reactions that could display such behavior are characterized by a strong van der Waals complex in the entrance valley, similarly to the gas-phase tunneling reactions described above, while the PES should also present a particularly low activation barrier to product formation. A recent theoretical study[12] of the reaction between the methylidyne radical (CH) and carbon dioxide ($CO_2$) suggests that this system could fulfil these requirements. The energy profile exhibited in Figure 1 indicates that the preferential pathway for the CH + $CO_2$ reaction follows the initial formation of a pre-reactive CH...$CO_2$ complex (stabilized by 8.8 kJ/mol with respect to reagents), which is converted to a stable intermediate molecular adduct (IM1) by means of a transition state structure (TSR1); a process presenting an activation energy of only 2.7 kJ/mol over the 300 – 700 K range. Through simulations of the species concentrations as a function of time, Vichietti *et al.*[12] showed that the elementary CH + $CO_2$ → IM1 step is rate determining for the global reaction, CH + $CO_2$ → products, over a broad temperature range (200 - 3500 K). Earlier experimental kinetic studies by Berman *et al.*[13] over the 297 - 676 K range support this theoretical hypothesis, finding an Arrhenius' activation energy of 2.9 ± 0.5 kJ/mol.

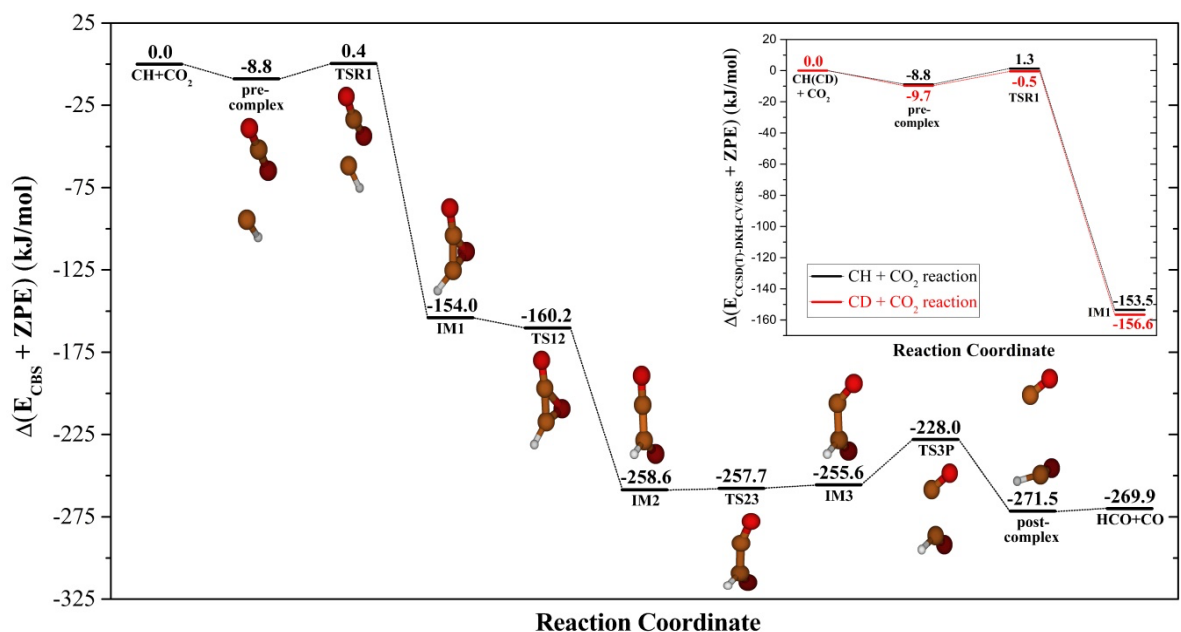

**Figure 1.** Energy profile (electronic energies extrapolated to the basis set limit, $E_{CBS}$, plus vibrational zero-point corrections, ZPE) relative to the reagents along the preferential pathway encountered by Vichietti et al.[12] for the CH + $CO_2$ → HCO + CO reaction, according to ROCCSD(T)/CBS//UCCSD/cc-pVDZ dual-level calculations (molecular structures were plotted with Molden 5.3).[14] The inset in the top right is the energy profile of the CH(CD) + $CO_2$ → IM1 elementary step obtained here in more advanced DKH-CV-CCSD(T)/CBS calculations.

Mechanistically, the rate determining step of the CH + $CO_2$ reaction (Figure 1) involves the approach of the CH radical to one of the C=O bonds of $CO_2$ (TSR1),[12] eventually forming the cyclic species (IM1). The C=O bond initially belonging to $CO_2$ then breaks, leading to the formation of an OC(H)-CO adduct (IM2) that is converted to another conformer (IM3) prior to product formation (HCO + CO). The essential point to retain here is that cleavage of the initial C-H bond is not predicted to occur, implying that eventual tunneling effects during C-H insertion might not be due to H-atom QMT.

To verify the possible influence of tunneling on this process, we report the results of a gas-phase kinetic study of the CH + $CO_2$ reaction below room temperature, where QMT effects are more likely to manifest. To further examine the role of QMT in the overall mechanism, rate constants for the CD + $CO_2$ reaction were also measured over the same range. Earlier kinetic measurements of these two processes have already shown that their reactivity is similar at room temperature.[15] Measurements were performed using the CRESU technique (Cinétique de Réaction en Écoulement Supersonique Uniforme or reaction kinetics in a uniform supersonic flow)[16] to generate the prerequisite low temperature gas flows. In addition to the experimental work, new rate constant calculations are also reported for the rate-determining step of the reactions under study, extending the preliminary work of Vichietti et al.[12]

## 2 Theoretical Methods

The thermochemical properties were calculated by using the equilibrium structures of stable points (reactants, product, saddle point and pre-complex) found previously by Vichietti et al.[12] for the rate-determining step of the global CH + $CO_2$ → products reaction (the CH + $CO_2$ → IM1 elementary step illustrated in Figure 1), which were obtained with unrestricted coupled cluster theory allowing single and double excitations (UCCSD)[17] in combination with the cc-pVDZ basis set.[18] The same geometries are also considered for the CD + $CO_2$ → IM1 process. In addition, the previous study[12] already demonstrated that the single-reference treatment devised here is reliable for these reactions. For example, the largest $T_1$ diagnostic value found for the structures investigated here was only 0.022 (for TSR1 and IM1), which is within the safe limits recommended in the literature for radicals, 0.045.[19]

Single-point calculations were performed at these stationary point structures using the restricted coupled cluster theory allowing single, double and connected triple excitations, CCSD(T).[20] Two more contributions were considered in this study to improve the description of thermochemical properties with respect to the previous work of Vichietti et al.:[12] (i) scalar relativistic effects by means of the second-order Douglas-Kroll-Hess Hamiltonian (DKH);[21-26] (ii) the contribution of core electron correlation (also known as core-valence correlation, CV). This combined method is labelled here as DKH-CV-CCSD(T). Therefore, appropriate basis sets for the DKH Hamiltonian including core-valence correlating functions[27] of quadruple- and quintuple-zeta qualities were employed, cc-pCVXZ-DK (X = Q,5),[28,29] and the results were extrapolated to the complete basis set (CBS) limit.[30] These calculations were performed by using the Molpro 2015 package.[31] It should be emphasized that the largest $T_1$ diagnostic value obtained in these DKH-CV-CCSD(T) calculations is only 0.020 (IM1). The thermochemical properties determined also considered the harmonic vibrational frequencies previously calculated with the UCCSD/cc-pVDZ approach scaled by a factor of 0.947[32] to take into account anharmonicity effects.

The high pressure limit thermal rate constants were obtained by transition state theory (TST) within the improved canonical variational theory (ICVT),[33] which maximizes the free energy along the reactional path. The minimum energy path was built by using a dual-level methodology, employing the DKH-CV-CCSD(T)/CBS//UCCSD/cc-pVDZ approach as the high-level choice and the UCCSD/cc-pVDZ method as the low-level one. This procedure is usually referred to in the literature as variational transition state theory with interpolated single-point energies (VTST-ISPE).[34] Tunneling effects were considered by the small curvature tunneling (SCT) method.[35] The chemical kinetic calculations were carried out by using the Polyrate 2008 package[36] and the Gaussrate 2009[37] interface between Gaussian 09[38] and Polyrate. The reaction rate calculations also considered the spin-orbit coupling of the CH radical by means of the same approach followed in Vichietti et al.[12] The reaction was considered as a one-step process and the pre-complex was included in the calculation of rate constants, as implemented in the Polyrate package. The derived rate constants considering the DKH-CV-CCSD(T)/CBS electronic structure results are listed in Tables S1 and S2 of the Supporting Information. The thermochemical properties of the CH(CD) + $CO_2$ → IM1 elementary step are presented in Table S3.

## 3 Experimental Methods

The CRESU technique was used for the present study. As the method has been previously described,[39,40] only the experimental details specific to the current investigation will be outlined here. Three Laval nozzles were employed during this study producing supersonic flows with temperatures of 177 K, 127 K, 75 K and 50 K (the same nozzle was used to generate flows at 177 K and 127 K using $N_2$ and Ar as carrier gases respectively). The characteristic properties of these flows (such as the temperature, velocity and density), are summarized in Table S4 alongside other relevant information.

These values were calculated from earlier measurements of the impact pressure as a function of distance from the nozzle using a Pitot tube and the stagnation pressure within the reservoir. To extend the kinetic measurements to higher temperature, measurements were also performed at room temperature (296 K) by removing the Laval nozzle. In this case, the flow velocity was significantly reduced, effectively using the reactor as a slow-flow flash photolysis apparatus.

CH($X^2\Pi_r$) radicals were generated by two methods during this study. The first employed the multiphoton dissociation of $CHBr_3$ molecules at 266 nm with ∼ 23 mJ of pulse energy. The photolysis laser was aligned along the supersonic flow, creating a column of CH radicals of uniform density. $CHBr_3$ was entrained in the flow by bubbling a small flow of carrier gas through liquid $CHBr_3$ held at a known pressure. An upper limit of $1 \times 10^{13}$ cm$^{-3}$ was estimated for the gas-phase concentration of $CHBr_3$ in the experiments from its saturated vapour pressure. The second employed the multiphoton dissociation of $CBr_4$ molecules at 266 nm, allowing C($^1D$) atoms to be created in the supersonic flow. $H_2$ (or $D_2$) molecules were added to the supersonic flow in excess concentrations (greater than $4.2 \times 10^{14}$ cm$^{-3}$), allowing CH (CD) radicals to be generated through the fast C($^1D$) + $H_2$ ($D_2$) reaction.[41,42] $CBr_4$ molecules were introduced into the flow by passing a small amount of Ar carrier gas over solid $CBr_4$ held at a known pressure. An upper limit of $2.5 \times 10^{13}$ cm$^{-3}$ was estimated for the gas-phase concentration of $CBr_4$ in these experiments. The C($^3P$) atoms generated by $CBr_4$ photolysis are unreactive with $H_2$ ($D_2$) and $CO_2$, so these atoms do not interfere with the kinetics of the CH + $CO_2$ reaction. This method allowed us to overcome issues related to cluster formation at 50 K. Indeed, as a large molecule with a non-zero electric dipole moment of 0.99 Debye, $CHBr_3$ readily formed clusters with coreagent $CO_2$ at this temperature, limiting the exploitable range of $CO_2$ concentrations. In contrast, the use of non-polar $CBr_4$ molecules allowed somewhat larger $CO_2$ concentrations to be used, leading to a greater range of pseudo-first-order rate constants and a more reliable measurement of the second-order rate constant. To ensure that the rate constants obtained with this method were not affected by secondary reactions, experiments were also performed using $CBr_4$ at higher temperatures (127 K and 296 K) where the results could be compared directly with those obtained by experiments employing $CHBr_3$ as the CH source.

CH($X^2\Pi_r$) radicals were followed by a chemiluminescence detection method during experiments. Here, molecular oxygen $O_2$ was added to the reactor so that the CH + $O_2$ → OH + CO reaction occurred, producing electronically excited OH($A^2\Sigma^+$) (hereafter denoted OH*). The fluorescence emission from OH* was detected using a UV sensitive photomultiplier tube and an interference filter centered on the OH ($A^2\Sigma^+$ → $X^2\Pi$) transition around 310 nm, in a similar manner earlier measurements of the CH + $H_2O$ reaction by Hickson et al.[43] A constant $O_2$ concentration used for any single series of measurements, with a maximum value of 2.3 × $10^{14}$ cm$^{-3}$. For most experiments, including all those employing $CHBr_3$ as the CH source, the $O_2$ concentration was much lower. This detection method provided certain advantages over the laser induced fluorescence (LIF) method usually employed during this type of experiment. Firstly, by detecting OH*, the entire CH temporal profile was traced for each photolysis laser shot, drastically reducing the time required for signal acquisition. Secondly, the simultaneous acquisition of all datapoints after each laser shot significantly lowered the potential for signal drift as a function of experiment time, thereby improving the accuracy of the derived decays when compared to the LIF method. The time dependent chemiluminescence signals consisted of 500 time points acquired by a 500 MHz digital oscilloscope for a range of excess $CO_2$ concentrations. The signal was averaged over 768 photolysis laser shots for each kinetic decay, while several kinetic decays were recorded for each value of the $CO_2$ concentration. At least eight $CO_2$ concentrations were used at each temperature. The chemiluminescence intensities were recorded at a fixed distance from the Laval nozzle. This distance corresponded to the maximum displacement from the nozzle where optimal flow conditions could be guaranteed, so that the decays could be exploited over as large a time period as possible. These values were determined during the supersonic flow characterization experiments described above.

All gases (Linde: Ar 99.999%, $CO_2$ 99.998%; Air Liquide: $N_2$ 99.999%, $H_2$ 99.999%, $D_2$ 99.8%, $CH_4$ 99.9995%) were flowed directly from cylinders with no further purification. The carrier gas and precursor flows were regulated by digital mass flow controllers which were calibrated using the pressure rise at constant volume method for the specific gas used.

**4 Results and Discussion**

The pseudo-first-order approximation was applied for all the experiments performed here, with a large excess concentration of $CO_2$ with respect to the CH(CD) concentration. Under these conditions, the CH(CD) radical signal (as traced by OH*(OD*) chemiluminescent emission) followed a simple exponential decay profile. Examples of the CH decay profiles recorded at 127 K are shown in Figure 2.

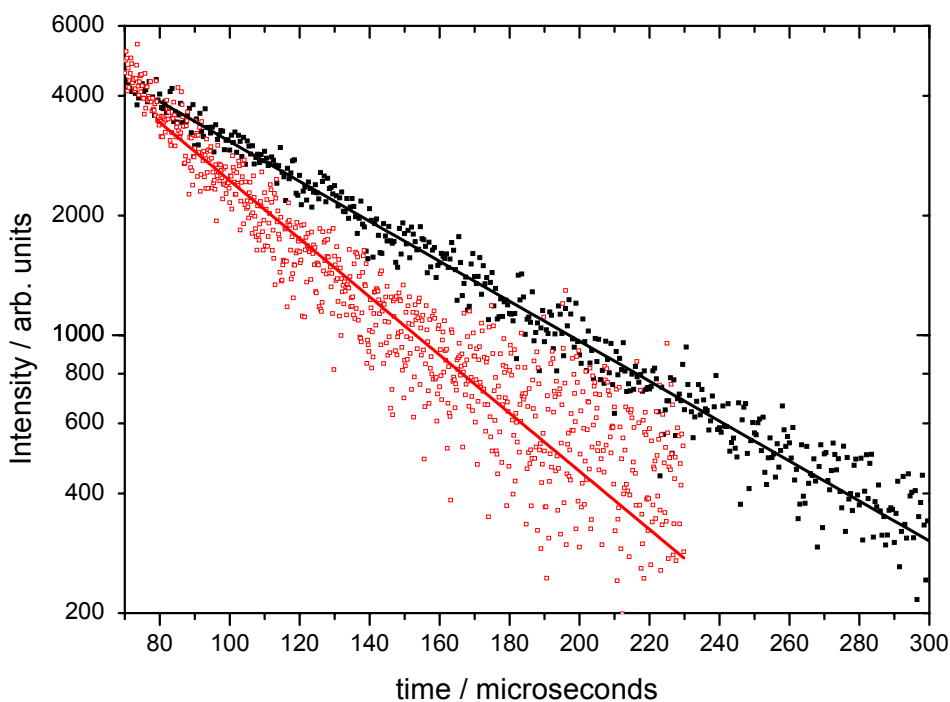

**Figure 2** CH(OH*) chemiluminescence signal as a function of time recorded at 127 K using the C($^1$D) +$H_2$ reaction as the source of CH radicals. [$H_2$] = 7.9 × $10^{14}$ cm$^{-3}$, [$O_2$] = 9.9 × $10^{13}$ cm$^{-3}$. (Red open squares) with [$CO_2$] = 6.1 × $10^{15}$ cm$^{-3}$ (black solid squares) measured signal in the absence of $CO_2$. Solid lines represent exponential fits to the individual datasets.

Here, the signal intensities are plotted on a logarithmic scale for clarity. Upon addition of excess $CO_2$ to the reactor, a clear enhancement is observed in the loss rate of CH or CD radicals, thereby allowing pseudo-first-order rate constants to be derived over a range of $CO_2$ concentrations. Exponential fits to these and other decays allowed the pseudo-first-order rate constants, $k_{1st} = k_{CH+CHBr_3}[CHBr_3] + k_{CH+O_2}[O_2] + k_{CH+CO_2}[CO_2] + k_L$, to be derived for experiments employing CHBr$_3$ as the CH precursor. Here $k_{CH+CHBr_3}$, $k_{CH+O_2}$ and $k_{CH+CO_2}$ are the second-order rater constants for the reaction of CH with CHBr$_3$, $O_2$ and $CO_2$ respectively and $k_\text{L}$ is an term representing additional first-order CH losses such as diffusion and reaction with impurities. For experiments employing CBr$_4$ as the CH(CD) precursor, the expression becomes $k_\text{1st} = k_\text{CH(CD)+CBr}_4[\text{CBr}_4] + k_{\text{CH+H}_2+\text{M}}[\text{H}_2][\text{Ar}] + k_{\text{CH(CD)+O}_2}[\text{O}_2] + k_{\text{CH(CD)+CO}_2}[\text{CO}_2] + k_\text{L}$.

In the absence of CO$_2$, CH(CD) radicals are lost by reaction with the precursor molecule (CHBr$_3$ or CBr$_4$), with O$_2$ (allowing CH(CD) radicals to be followed through OH chemiluminescence) and by diffusion/reaction with impurities. Additionally, in the case of experiments with CBr$_4$ and H$_2$ we also need to consider the CH + H$_2$ termolecular association reaction. As the precursor (CHBr$_3$ or CBr$_4$) concentration, [O$_2$] and [H$_2$] are constant for any single series of experiments, the observed decay rate increase when CO$_2$ is added to the flow originates solely from the additional contribution of the CH(CD) + CO$_2$ reaction. Second-order rate constants, $k_{\text{CH(CD)+CO}_2}$, were thus extracted through weighted linear least-squares fits of $k_\text{1st}$ as a function of [CO$_2$] for data obtained at a specified temperature. Examples of such fits for both the CH + CO$_2$ and CD + CO$_2$ reactions are shown in Figure 3.

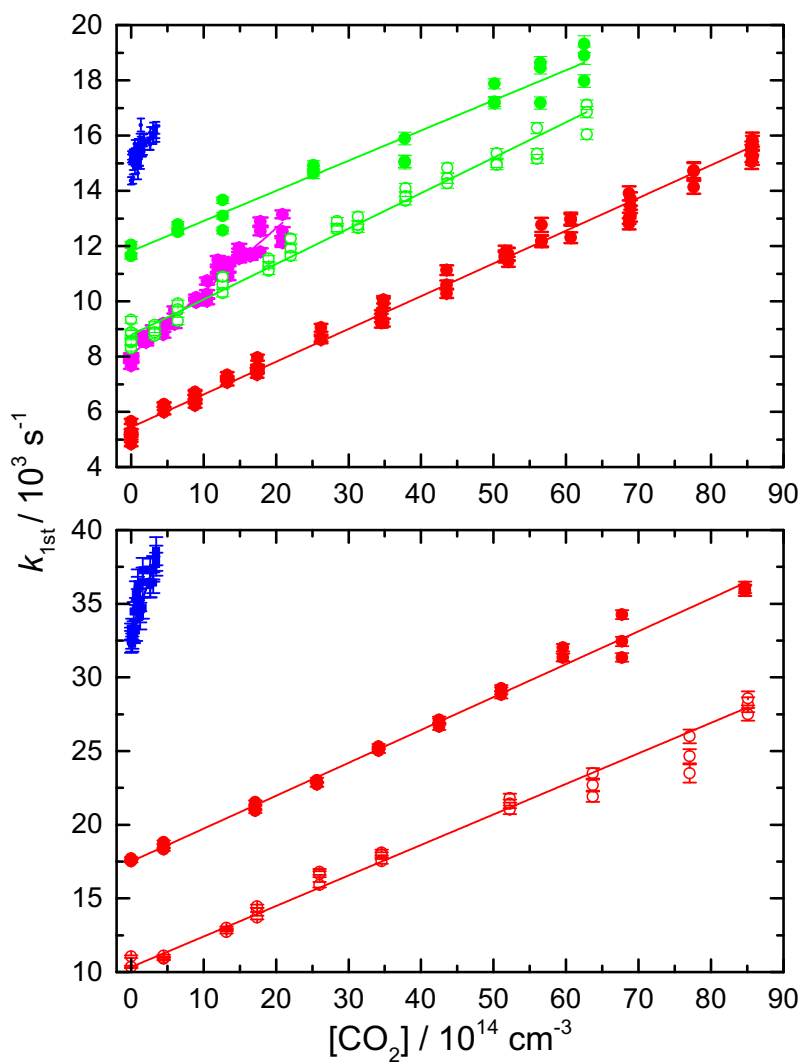

**Figure 3.** Derived pseudo-first-order rate constants $k_{1st}$ as a function of [$CO_2$] for a range of temperatures. (Upper panel) The CH + $CO_2$ reaction (red solid circles) 296 K data with $CHBr_3$; (green open circles) 127 K data with $CHBr_3$; (green solid circles) 127 K data with $CBr_4$ and [$H_2$] = 7.9 × $10^{14}$ cm$^{-3}$; (magenta solid circles) 75 K data with $CHBr_3$; (blue solid circles) 50 K data with $CBr_4$ and [$H_2$] = 4.2 × $10^{14}$ cm$^{-3}$. (Lower panel) The CD + $CO_2$ reaction (red solid circles) 296 K data with $CBr_4$ and [$D_2$] = 1.5 × $10^{15}$ cm$^{-3}$; (red open circles) 296 K data with $CBr_4$ and [$D_2$] = 7.5 × $10^{14}$ cm$^{-3}$; (blue solid circles) 50 K data with $CBr_4$ and [$D_2$] = 4.2 × $10^{14}$ cm$^{-3}$.

As it was only possible to measure rate constants at 50 K using $CBr_4$ and $H_2$($D_2$) as the CH(CD) radical source (the formation of clusters between $CO_2$ and $CHBr_3$ severely limited the useful range of [$CO_2$] when $CHBr_3$ was used as the CH source at this temperature), test experiments were performed at 127 K and 296 K to check that the two methods yielded similar results.

At 127 K, using $CHBr_3$ as the CH source, a value of $(1.28 \pm 0.13) \times 10^{-12}$ cm$^3$ s$^{-1}$ was obtained for the second-order rate constant which was close to the value obtained by the method using $CBr_4$ and $H_2$ of $(1.09 \pm 0.12) \times 10^{-12}$ cm$^3$ s$^{-1}$. At 296 K, the measured rate constant values using $CHBr_3$ to produce CH were $(1.19 \pm 0.12) \times 10^{-12}$ cm$^3$ s$^{-1}$ and $(1.35 \pm 0.14) \times 10^{-12}$ cm$^3$ s$^{-1}$ respectively for Ar and $N_2$ as the carrier gases respectively. The method using $CBr_4$ and $H_2$ to produce CH radicals gave a slightly higher value of $(1.71 \pm 0.17) \times 10^{-12}$ cm$^3$ s$^{-1}$ (all measurements with $CBr_4$ and $H_2$ as the CH source were performed with Ar as the carrier gas due to the rapid quenching of C($^1$D) atoms by $N_2$). Similarly, two experiments were performed at 296 K for the CD + $CO_2$ reaction employing different $D_2$ concentrations, allowing us to check for possible interferences from secondary reactions. Rate constant values of $(2.23 \pm 0.23) \times 10^{-12}$ cm$^3$ s$^{-1}$ and $(2.07 \pm 0.21) \times 10^{-12}$ cm$^3$ s$^{-1}$ were derived using [$D_2$] = $1.49 \times 10^{15}$ cm$^{-3}$ and $7.46 \times 10^{14}$ cm$^{-3}$ respectively.

**Effects of the CH+$O_2$ tracer reaction and other secondary reactions**

The photolysis of $CHBr_3$ at 266 nm could lead to the formation of some excited state CH($a^4\Sigma$) radicals in our experiments which could interfere with the kinetics of the CH($X^2\Pi_r$) + $CO_2$ reaction. It has been estimated[44] that the photolysis of $CHBr_3$ at 248 nm produces five times more ground state CH($X^2\Pi_r$) radicals than electronically excited CH($a^4\Sigma$) ones, although the relative yield of CH($a^4\Sigma$) radicals is likely to be even lower at 266 nm. Indeed OH* is also produced by the CH($a^4\Sigma$) + $O_2$ reaction with a room temperature rate constant of $2.6 \times 10^{-11}$ cm$^3$ s$^{-1}$.[45] Previous work by Hou and Bayes has shown that CH($a^4\Sigma$) radicals react slowly with $CO_2$, with a room temperature rate constant of less than $3 \times 10^{-13}$ cm$^3$ s$^{-1}$.[46] The intercepts of second-order plots such as those shown in Figure 3 represent the sum of several contributions as described above. In order to evaluate our understanding of the overall chemistry occurring in the reactor, we can consider the measurements performed for the CH + $CO_2$ reaction at 50 K, where the derived intercept value is approximately 15000 s$^{-1}$. As the chemiluminescent OH* emission of the CH($X^2\Pi_r$) + $O_2$ reaction was used to follow CH($X^2\Pi_r$) radicals, a small fraction of CH radicals is consumed by this process. The temperature dependent rate constants for the CH + $O_2$ reaction have been measured previously[47] allowing us to estimate a pseudo-first-order loss rate for CH of ~ 6000 s$^{-1}$ at 50 K. Additionally, using the rate constants for the CH + $H_2$ termolecular association reaction derived by Brownsword et al.[48] leads to an estimate of the pseudo-first-order loss rate for CH of ~ 400 s$^{-1}$ by this process at 50 K. If we assume a large rate constant for the reaction of CH radicals with the precursor molecules $CHBr_3$ or $CBr_4$ (corresponding to reaction occurring for essentially every collision), we can estimate a contribution for this process to the CH pseudo-first-order loss rate of a few thousand s$^{-1}$ at 50 K. Finally, diffusional losses also amount to a few thousand s$^{-1}$ at this temperature based on earlier work. As the sum of these contributions is close to the measured intercept value, it does not appear that any major reactive processes are unaccounted for.

In previous work on the CH + $H_2O$ reaction, Hickson et al.[43] examined the influence of the $O_2$ concentration on the derived rate constants for this reaction. Test experiments were performed at 296 K with different fixed $O_2$ concentrations, each over a range of coreagent concentrations. These measurements yielded second-order plots with the same slopes, indicating that the tracer reaction products did not interfere with the CH kinetics.

**Product *versus* complex formation in the CH + $CO_2$ reaction**

To check that the recorded CH decays in the presence of $CO_2$ resulted in real product formation (rather than just simply complex formation with a long enough lifetime for it to remove CH from the flow on the timescale of the experiment), additional experiments were performed at 127 K. As can be seen from Figure 1, the products of the CH + $CO_2$ reaction are considered to be HCO + CO, while a significant fraction of the exothermic energy released by the reaction is expected to be carried away by the HCO fragment. Under these conditions, the HCO radical could further dissociate to H + CO. Here we detected the formation of atomic hydrogen by laser induced fluorescence at 121.567 nm in a similar manner to our previous work on the C + $H_2O$ reaction.[2] These experiments employed $CHBr_3$ photolysis at 266 nm as the source of CH radicals. In order to calibrate the H-atom production efficiency of the CH + $CO_2$ reaction to obtain quantitative information, the H-atom yield was compared with that of a reference process, namely the CH + $CH_4$ reaction which forms $C_2H_4$ + H as the exclusive products.[49] Figure 4 shows the results of one of the six experiments that were performed.

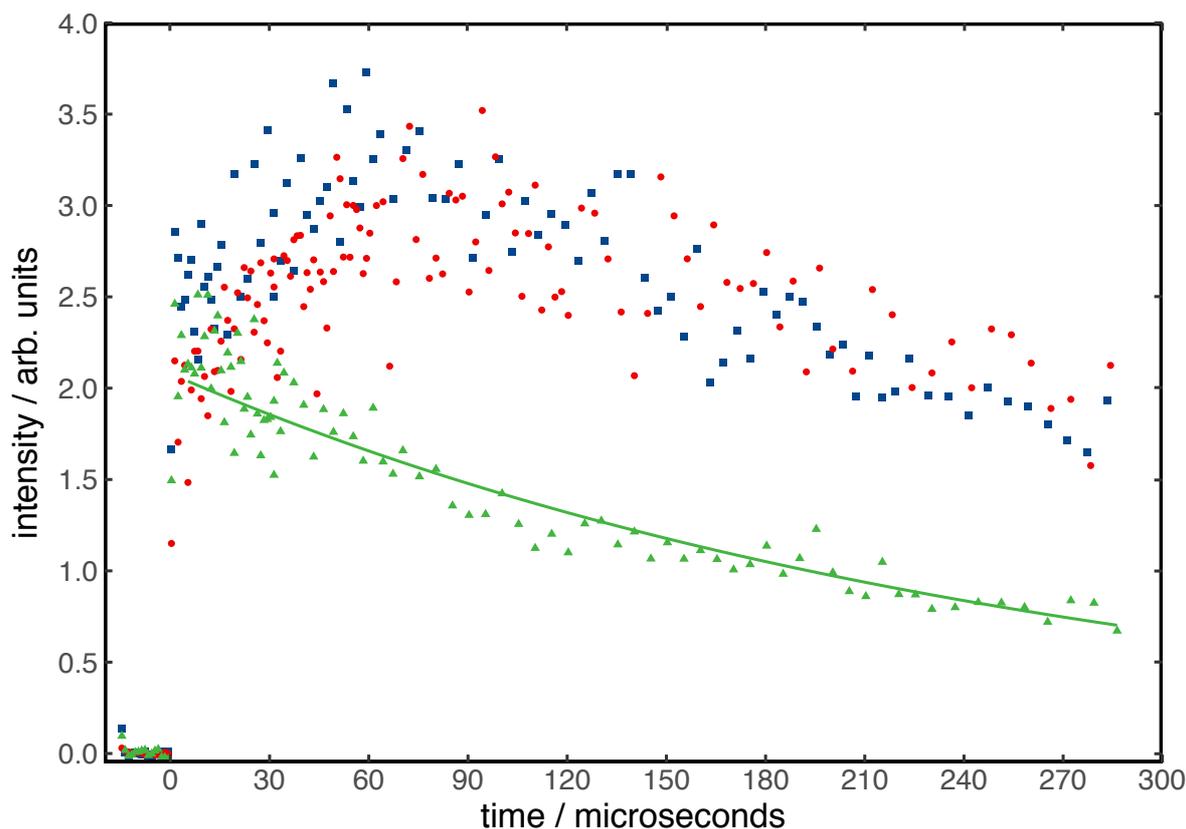

**Figure 4.** H-atom temporal profiles as a function of time recorded at 127 K. (Red solid circles) H-atom signal from the CH + CO$_2$ reaction with [CO$_2$] = 6.2 × 10$^{15}$ cm$^{-3}$. (Blue solid squares) H-atom signal from the CH + CH$_4$ reaction with [CH$_4$] = 4.1 × 10$^{13}$ cm$^{-3}$. (Green solid triangles) H-atom signal recorded in the absence of CH$_4$ and CO$_2$, corresponding to nascent H-atoms produced by CHBr$_3$ photolysis. The green solid line represents a single exponential fit to the data.

Figure 4 clearly shows that the photolysis of CHBr$_3$ at 266 nm also results in the direct production of H-atoms, which decay slowly through diffusion from the observation region (green datapoints). This signal can therefore be considered as the baseline level for subsequent H-atom yield measurements. Here, CO$_2$ and CH$_4$ concentrations were carefully chosen to provide identical first-order production rates (~8000 s$^{-1}$) for H-atoms (so that H-atom diffusional losses were identical), based on the measured rate constants for these processes. When CO$_2$ is added to the flow a clear production of H-atoms is observed (red datapoints), over and above the baseline level, followed by a slow decay of these atoms. When CH$_4$ is added to the reactor (in the absence of CO$_2$), a similar H-atom biexponential type profile is obtained (blue datapoints) to the one recorded for the CH + CO$_2$ reaction with yields that are clearly very similar in magnitude to those of the target reaction. To obtain quantitative yields from these data would require a subtraction of the nascent H-atom signal intensity from the signal obtained from the target and reference reactions. Unfortunately, this analysis resulted in large errors, so that quantitative yields could not be extracted. Nevertheless, the large observed H-atom yields from the CH + $CO_2$ reaction allow us to draw a couple of important qualitative conclusions: (1) the HCO product of the CH + $CO_2$ reaction dissociates almost entirely to H + CO in the present experiments (2) complex formation represents only a very small fraction of the overall end product of the CH + $CO_2$ reaction, with the majority of successful collisions resulting in real product formation.

The measured rate constants are listed in Table 1 and are shown in Figure 5 alongside the present theoretical results and previous data from the literature.

**Table 1** Measured rate constants for the CH + $CO_2$ and CD + $CO_2$ reactions

| T / K | [$CO_2$] / $10^{14}$ cm$^{-3}$ | $k_{CH+CO_2}$ (CHBr$_3$) / $10^{-12}$ cm$^3$ s$^{-1}$ | [$CO_2$] / $10^{14}$ cm$^{-3}$ | $k_{CH+CO_2}$ (CBr$_4$) / $10^{-12}$ cm$^3$ s$^{-1}$ | [$CO_2$] / $10^{14}$ cm$^{-3}$ | $k_{CD+CO_2}$ (CBr$_4$) / $10^{-12}$ cm$^3$ s$^{-1}$ |
|---|---|---|---|---|---|---|
| 296 | 0 - 85.7[a] | (1.1 ± 0.12)[c] *63*[d] | 0 - 85.4 | (1.71 ± 0.17) *30* | 0 - 84.7 | (2.23 ± 0.23)[e] *30* |
| 296 | 0 - 85.7[b] | (1.35 ± 0.14) *27* | | | 0 - 85.1 | (2.07 ± 0.21)[f] *30* |
| 177 | 0 - 45.9 | (1.15 ± 0.13) *42* | | | | |
| 127 | 0 - 62.9 | (1.28 ± 0.13) *42* | 0 - 62.5 | (1.09 ± 0.12) *24* | | |
| 75 | 0 - 21.4 | (2.25 ± 0.23) *46* | | | | |
| 50 | | | 0 - 3.5 | (4.02 ± 0.68) *51* | 0 - 3.3 | (14.85 ± 2.0) *30* |

[a]Experiments conducted with Ar as the carrier gas. [b]Experiments conducted with $N_2$ as the carrier gas. [c]Errors are cited at the level of a single standard deviation from the mean and comprise an additional 10 % contribution from possible systematic errors. [d]Figures in italics represent the number of individual experiments. [e][$D_2$] = 1.49 × $10^{15}$ cm$^{-3}$. [f][$D_2$] = 7.46 × $10^{14}$ cm$^{-3}$.

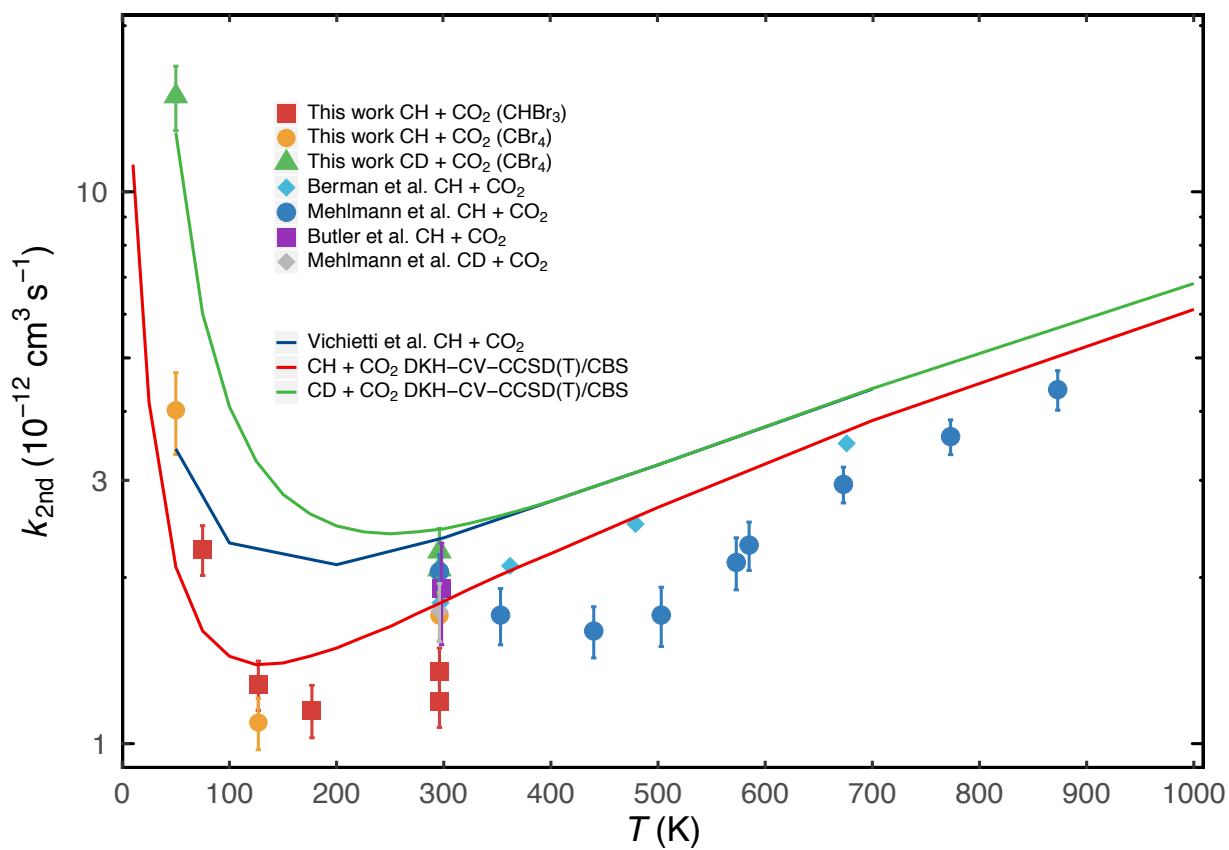

**Figure 5.** Temperature dependence of the rate constants for the CH (CD) + $CO_2$ reactions. **The CH + $CO_2$ reaction**. Experimental studies; (purple square) Butler et al.[50]; (cyan diamonds) Berman et al.[13]; (Blue circles) Mehlmann et al.[15]; (red squares) this work - $CHBr_3$ as CH precursor; (orange circles) this work - $CBr_4$ and $H_2$ precursors. Theoretical studies; (blue line) Vichietti et al.[12]; (red line) this work - ICVT/SCT calculations at the DKH-CV-CCSD(T)/CBS level. **The CD + $CO_2$ reaction**. Experimental studies; (gray diamond) Mehlmann et al.[15]; (green triangles) this work - $CBr_4$ and $H_2$ as CH precursors. Theoretical studies; (green solid line) this work - ICVT/SCT calculations at the DKH-CV-CCSD(T)/CBS level.

It can be seen from Figure 5 that the measured room temperature rate constants of (1.19-1.71) × $10^{-12}$ $cm^3$ $s^{-1}$ for the CH + $CO_2$ reaction agree well with those obtained in previous work,[13, 15, 50] (1.80-2.05) × $10^{-12}$ $cm^3$ $s^{-1}$, thereby validating the two experimental methods employed here. The experimental reaction rates present small variations between 100 K and 300 K, between 1.1 and 1.3 × $10^{-12}$ $cm^3$ $s^{-1}$. However, the CH + $CO_2$ reaction becomes much faster below 100 K and the rate constant reaches a value of (4.02 ± 0.68) × $10^{-12}$ $cm^3$ $s^{-1}$ at 50 K.

Figure 5 also shows that, while the previous rate constant calculations of Vichietti et al.[12] already capture the main features of the upturn at low temperatures, a factor of two discrepancy is observed between theory and experiment in the intermediate plateau region. However, as CV correlation and scalar relativistic effects are accounted for in the present electronic structure calculations, the respective ICVT/SCT rate constants provide much better agreement with the experimental results. Moreover, this study clearly demonstrates that the proposed tunneling mechanism involving CH insertion can adequately explain the experimental measurements since the quantum tunneling correction (SCT) is the primary factor responsible for the reaction rate increases as the temperature is lowered (see Table S1 and Fig. S1). In other words, the rate constants for the CH + $CO_2$ reaction obtained without quantum tunneling corrections would erroneously decrease as the temperature is lowered, while good agreement with experimental observations occurs only through the inclusion of such corrections. In addition, since the reaction coordinate for the rate-determining step involves the displacement of the whole CH fragment, this suggests the involvement of a heavy-particle tunneling mechanism.

As a rigorous test of our understanding of the reaction mechanism, we also performed kinetic measurements and calculations for the CD + $CO_2$ system. Indeed, if H-atom tunneling was responsible for the observed acceleration at low temperature in the CH + $CO_2$ reaction then we would expect that the rate increase should be less pronounced for the deuterated equivalent due to the lower tunneling probability of D versus H. Surprisingly, the calculated and measured rate constants shown in Figure 3 present the opposite behavior, both displaying large increases below room temperature with almost perfect agreement between theory and experiment. A closer look at the electronic structure calculations (see Figure 1 and Table S3 in the Supporting Information) tells us that this effect originates from zero point energy (ZPE) contributions. Hence, while the ZPE corrected adiabatic barrier height ($\Delta V_a^{G,‡}$ in Table S3) of the rate-determining step is positive with respect to CH and $CO_2$ (1.3 kJ/mol), the larger reduced mass of the equivalent deuterated species lowers $\Delta V_a^{G,‡}$ to such an extent that it falls below the reagent asymptote (-0.5 kJ/mol), becoming a reef structure rather than a real barrier (see also the top right inset in Figure 1). Although this finding has little effect on room temperature results of both reactional systems, as back dissociation of the CH…$CO_2$ and CD…$CO_2$ pre-reactive complexes dominates (both the present and previous experimental studies[15] of the CD + $CO_2$ reaction derived reaction rates that were similar in magnitude to the CH + CO$_2$ ones), the passage forward over this reef structure is favored instead at lower temperature, leading to a large enhancement of the reaction rate. It is worth noting that tunneling from states with energies that lie below the reactants on the minimum energy path was not considered since the reaction was treated as a one-step process. Non-classical reflection was not taken into account. However, we believe that these contributions should be small. Moreover, some error cancellation is expected from both contributions. In summary, the rate constant increases observed for the CD + CO$_2$ reaction as temperatures are lowered are due to ZPE corrections, resulting in a submerged barrier with respect to reactants. Consequently, the CD + CO$_2$ reaction is comparable to a number of other systems presenting similar features over the PES leading from reagents to products, such as the CH + H$_2$O reaction[43] as well as the reactions of C($^3$P) with NH$_3$[51,52] and CH$_3$OH.[53]

## 5 Astrochemical Interest

The reactions of CH and CD with CO$_2$ could also be interesting from an astrochemical perspective since related models predict that this species should be present at high relative abundance levels (10$^{-7}$ with respect to H$_2$), although gas-phase CO$_2$ has never been detected in the interstellar medium due to the absence of a permanent electric dipole moment. As the main source of protonated carbon dioxide (HOCO$^+$) is thought to be the H$_3^+$ + CO$_2$ reaction,[54] the predicted CO$_2$ abundance has been indirectly confirmed through observations of HOCO$^+$. Given the predicted rate of the CH + CO$_2$ reaction of $1.1 \times 10^{-11}$ cm$^3$ s$^{-1}$ at 10 K, this reaction is likely to play only a minor role in the loss of interstellar CH, although it could be an important source of interstellar HCO (formyl) radicals. Furthermore, the large measured difference in the reactivity between CH or CD with CO$_2$ at low temperatures may represent an important mechanism for deuterium enrichment of formyl radicals in the gas-phase interstellar medium. Indeed, the present calculations predict rate constants of $4.2 \times 10^{-12}$ cm$^3$ s$^{-1}$ and $9.5 \times 10^{-11}$ cm$^3$ s$^{-1}$ for the CH + CO$_2$ and CD + CO$_2$ reactions, respectively, at 25 K; resulting in a difference of more than a factor of twenty in reactivity which could lead to a significantly enhanced gas-phase DCO/HCO ratio.

This work could also have implications for the chemistry of interstellar ices where CO$_2$ is thought to be present at high levels with respect to water ice (10-50%).[55] In these environments, an HCOCO adduct is generated by the addition of a physisorbed CH radical to solid CO$_2$, providing a potential mechanism for glycol CHOCHO formation through H addition (which are mobile at 10 K) to HCOCO. Another potential route for HCOCO formation is through the formation of an initial complex between atomic carbon and $CO_2$ ice (C...$CO_2$),[56] as C does not react with $CO_2$ directly. Instead, a hydrogen atom scans the surface until it encounters the C...$CO_2$ complex, reacting to produce CH...$CO_2$ before going on to form HCOCO. However, considering the very low binding energy of HCO-CO,[12] the main products of the CH + $CO_2$ reaction are likely to be HCO + CO or H + CO + CO as the available excess energy to be redistributed to the ice phonon modes is probably too great to effectively stabilize HCOCO. This process could instead represent an important mechanism for the reconversion of $CO_2$ to CO on interstellar ices.

# 6 Conclusions

Here we present the results of a joint experimental and theoretical study of the CH(CD) + $CO_2$ reaction at low temperature. On the experimental side, the Laval nozzle (CRESU) method was employed, coupled with pulsed laser photolysis production of CH(CD), with these radicals being detected by chemiluminescence. On the theoretical side, thermal rate constants were obtained by transition state theory within the improved canonical variational theory including quantum tunneling corrections. These calculations were based on the thermochemical properties for stationary point structures of the CH(CD) + $CO_2$ system obtained with a high-level electronic correlation methodology including scalar relativistic and core-valence correlation effects. The rate of the CH + $CO_2$ reaction is seen to accelerate below 100 K, with good agreement between experiment and theory. Supplementary experiments detecting atomic hydrogen formed from product HCO dissociation clearly show that real product formation is the major outcome of this process. Mechanistically, the reaction is shown to occur by means of quantum mechanical tunneling of the CH radical through the activation barrier. To exclude possible contributions from H atom tunneling effects, measurements and calculations were also performed for the CD + $CO_2$ reaction. This deuterated process was seen to react much more rapidly at low temperature as zero point energy effects eliminate the barrier to product formation. The possible astrochemical implications of these reactions are discussed.

**Acknowledgements**

RKFS, RMV, FBCM, and RLAH gratefully acknowledge the financial assistance of The São Paulo Research Foundation (FAPESP) under grants 2014/23714-1, 2018/05691-5, 2019/25105-6, and 2019/07671-4 and National Council for Scientific and Technological Development (CNPq) under grants 301211/2018-3, 307136/2019-1, 404337/2016-3, 407760/2018-0 and 305788/2018-3. K. M. H. acknowledges support from the French program ''Physique et Chimie du Milieu Interstellaire'' (PCMI) of the CNRS/INSU with the INC/INP co-funded by the CEA and CNES as well as funding from the ''Program National de Planétologie'' (PNP) of the CNRS/INSU.

**References**

1. Shannon, R. J.; Blitz, M. A.; Goddard, A.; Heard, D. E. Accelerated Chemistry in the Reaction Between the Hydroxyl Radical and Methanol at Interstellar Temperatures Facilitated by Tunnelling. *Nat. Chem.* **2013,** *5* (9), 745-749.

2. Hickson, K. M.; Loison, J.-C.; Nuñez-Reyes, D.; Méreau, R. Quantum Tunneling Enhancement of the C + $H_2O$ and C + $D_2O$ Reactions at Low Temperature. *J. Phys. Chem. Lett.* **2016,** *7* (18), 3641-3646.

3  Shannon, R. J.; Caravan, R. L.; Blitz, M. A.; Heard, D. E. A Combined Experimental and Theoretical Study of Reactions between the Hydroxyl Radical and Oxygenated Hydrocarbons Relevant to Astrochemical Environments. *Phys. Chem. Chem. Phys.* **2014,** *16*, 3466-3478.

4. Smith, I. W. M.; Sage, A. M.; Donahue, N. M.; Herbst, E.; Quan, D. The Temperature-Dependence of Rapid Low Temperature Reactions: Experiment, Understanding and Prediction. *Faraday Discuss.* **2006,** *133*, 137.

5. Nguyen, T. L.; Xue, B. C.; Weston, R. E.; Barker, J. R.; Stanton, J. F. Reaction of HO with CO: Tunneling Is Indeed Important. *J. Phys. Chem. Lett.* **2012,** *3* (11), 1549-1553.

6. Tizniti, M.; Le Picard, S. D.; Lique, F.; Berteloite, C.; Canosa, A.; Alexander, M. H.; Sims, I. R. The Rate of the F + $H_2$ Reaction at Very Low Temperatures. *Nat. Chem.* **2014,** *6* (2), 141-145.

7. Liu, Y.; Song, H.; Xie, D.; Li, J.; Guo, H. Mode Specificity in the OH + $HO_2$ → $H_2O$ + $O_2$ Reaction:Enhancement of Reactivity by Exciting a Spectator Mode. J*. Am. Chem. Soc.* **2020**, *142*, 3331–3335.

8. Carpenter, B. K. Heavy-atom Tunneling as the Dominant Pathway in a Solution-phase Reaction? Bond Shift in Antiaromatic Annulenes. *J. Am. Chem. Soc.* **1983,** *105* (6), 1700-1701.

9. Zuev, P. S.; Sheridan, R. S.; Albu, T. V.; Truhlar, D. G.; Hrovat, D. A.; Borden, W. T. Carbon Tunneling from a Single Quantum State. *Science* **2003,** *299* (5608), 867.

10. Goldanskii, V. I.; Frank-Kamenetskii, M. D.; Barkalov, I. M. Quantum Low-Temperature Limit of a Chemical Reaction Rate. *Science* **1973,** *182* (4119), 1344.

11. Goldanskii, V. I. Chemical Reactions at Very Low Temperatures. *Annu. Rev. Phys. Chem.* **1976,** *27* (1), 85-126.

12. Vichietti, R. M.; Spada, R. F. K.; da Silva, A. B. F.; Machado, F. B. C.; Haiduke, R. L. A. A Proposal for the Mechanism of the CH + $CO_2$ Reaction. *ACS Omega* **2019,** *4* (18), 17843-17849.

13. Berman, M. R.; Fleming, J. W.; Harvey, A. B.; Lin, M. C. Temperature Dependence of CH Radical Reactions with $O_2$, NO, CO and $CO_2$. *Symp. Int. Combust. Proc.* **1982,** *19*, 73.

14. Schaftenaar, G.; Noordik, J. H. Molden: a Pre- and Post-processing Program for Molecular and Electronic Structures*. *J. Comput. Aided Mol. Des.* **2000,** *14* (2), 123-134.

15. Mehlmann, C.; Frost, M. J.; Heard, D. E.; Orr, B. J.; Nelson, P. F. Rate Constants for Removal of CH(D) (v=0 and 1) by Collisions with $N_2$, CO, $O_2$, NO and $NO_2$ at 298 K and with $CO_2$ at 296 $\leq$ T/K $\leq$ 873. *J. Chem. Soc. Faraday Trans.* **1996**, 2335 - 2341.

16. Rowe, B. R.; Dupeyrat, G.; Marquette, J. B.; Gaucherel, P. Study of the Reactions $N_2^+$ + $2N_2$ -> $N_4^+$ + $N_2$ and $O_2^+$ + $2O_2$ -> $O_4^+$ + $O_2$ from 20 to 160 K by the CRESU Technique. *J. Chem. Phys.* **1984,** *80* (10), 4915-4921.

17. Scuseria, G. E.; Schaefer, H. F. Is Coupled Cluster Singles and Doubles (CCSD) More Computationally Intensive than Quadratic Configuration Interaction (QCISD)? *J. Chem. Phys.* **1989,** *90* (7), 3700-3703.

18. Dunning Jr., T. H. Gaussian Basis Sets For Use in Correlated Molecular Calculations. I. The Atoms Boron Through Neon and Hydrogen. *J. Chem. Phys.* **1989,** *90*, 1007-1024.

19. Rienstra-Kiracofe, J. C.; Allen, W. D.; Schaefer, H. F.; The $C_2H_5$ + $O_2$ Reaction Mechanism: High-Level Ab Initio Characterizations. *J. Phys. Chem. A* **2000,** *104* (44), 9823-9840.

20. Hampel, C.; Peterson, K. A.; Werner, H.-J. A Comparison of the Efficiency and Accuracy of the Quadratic Configuration Interaction (QCISD), Coupled Cluster (CCSD), and Brueckner Coupled Cluster (BCCD) Methods. *Chem. Phys. Lett.* **1992,** *190* (1), 1-12.

21. Jansen, G.; Hess, B. A. Revision of the Douglas-Kroll Transformation. *Phys. Rev. A* **1989,** *39* (11), 6016-6017.

22. Hess, B. A. Relativistic Electronic-Structure Calculations Employing a Two-Component No-Pair Formalism with External-Field Projection Operators. *Phys. Rev. A* **1986,** *33* (6), 3742-3748.

23. Hess, B. A. Applicability of the No-Pair Equation with Free-Particle Projection Operators to Atomic and Molecular Structure Calculations. *Phys. Rev. A* **1985,** *32* (2), 756-763.

24. Douglas, M.; Kroll, N. M. Quantum Electrodynamical Corrections to the Fine Structure of Helium. *Ann. Phys.* **1974,** *82* (1), 89-155.

25. Reiher, M.; Wolf, A. Exact Decoupling of the Dirac Hamiltonian. I. General Theory. *J. Chem. Phys.* **2004,** *121* (5), 2037-2047.

26. Reiher, M.; Wolf, A. Exact Decoupling of the Dirac Hamiltonian. II. The Generalized Douglas–Kroll–Hess Transformation up to Arbitrary Order. *J. Chem. Phys.* **2004,** *121* (22), 10945-10956.

27. Peterson, K. A.; Dunning, T. H. Accurate Correlation Consistent Basis Sets for Molecular Core–valence Correlation Effects: The Second Row Atoms Al–Ar, and the First Row Atoms B–Ne Revisited. *J. Chem. Phys.* **2002,** *117* (23), 10548-10560.

28. Woon, D. E.; Dunning, T. H. Gaussian Basis Sets for use in Correlated Molecular Calculations. V. Core-valence Basis Sets for Boron through Neon. *J. Chem. Phys.* **1995,** *103* (11), 4572-4585.

29. de Jong, W. A.; Harrison, R. J.; Dixon, D. A. Parallel Douglas–Kroll Energy and Gradients in NWChem: Estimating Scalar Relativistic Effects using Douglas–Kroll Contracted Basis Sets. *J. Chem. Phys.* **2000,** *114* (1), 48-53.

30. Peterson, K. A.; Feller, D.; Dixon, D. A. Chemical Accuracy in Ab Initio Thermochemistry and Spectroscopy: Current Strategies and Future Challenges. *Theor. Chem. Acc.* **2012,** *131* (1), 1079.

31. Werner, H.-J.; Knowles, P. J.; Knizia, G.; Manby, F. R.; Schütz, M. Molpro: a General-Purpose Quantum Chemistry Program Package. *WIREs Comput. Mol. Sci.* **2012,** *2* (2), 242-253.

32. Johnson III, R. D. Computational Chemistry Comparison and Benchmark Database NIST Standard Reference Database, Number 101, Release 2015, 17b. http://cccbdb.nist.gov/.

33. Truhlar, D. G.; Garrett, B. C. Variational Transition-state Theory. *Acc. Chem. Res.* **1980,** *13* (12), 440-448.

34. Chuang, Y.-Y.; Corchado, J. C.; Truhlar, D. G. Mapped Interpolation Scheme for Single-Point Energy Corrections in Reaction Rate Calculations and a Critical Evaluation of Dual-Level Reaction Path Dynamics Methods. *J. Phys. Chem. A* **1999,** *103* (8), 1140-1149.

35. Liu, Y. P.; Lynch, G. C.; Truong, T. N.; Lu, D. H.; Truhlar, D. G.; Garrett, B. C. Molecular Modeling of the Kinetic Isotope Effect for the [1,5]-Sigmatropic Rearrangement of Cis-1,3-pentadiene. *J. Am. Chem. Soc.* **1993,** *115* (6), 2408-2415.

36. Zheng, J.; Zhang, S.; Lynch, B. J.; Corchado, J. C.; Chuang, Y.-Y.; Fast, P. L.; Hu, W.-P.; Liu, Y. P.; Lynch, G. C.; Nguyen, K. A.; et al. *Polyrate-version 2008* University of Minnesota, Minneapolis, MN, 2008, 2008.

37. Zheng, J.; Zhang, S.; Corchado, J. C.; Chuang, Y.-Y.; Coitiño, E. L.; Ellingson, B. A.; Truhlar, D. G. *GAUSSRATE – version 2009-A*, Department of Chemistry and Supercomputing Institute, University of Minnesota, Minneapolis, Minnesota 55455, 2009.
University of Minnesota, Minneapolis, Minnesota 55455, 2009.

38. R. A., Frisch, M. J.; Trucks, G. W.; Schlegel, H. B.; Scuseria, G. E.; Robb, M. A.; Cheeseman, J. R.; Scalmani, G.; Barone, V.; Mennucci, B.; Petersson, G. A.; et al. Gaussian 09, *Gaussian, Inc., Wallingford CT, 2009*.

39. Daugey, N.; Caubet, P.; Bergeat, A.; Costes, M.; Hickson, K. M. Reaction kinetics to low temperatures. Dicarbon + Acetylene, Methylacetylene, Allene and Propene from 77 ≤ T ≤ 296 K. *Phys. Chem. Chem. Phys.* **2008,** *10* (5), 729-737.

40. Daugey, N.; Caubet, P.; Retail, B.; Costes, M.; Bergeat, A.; Dorthe, G. Kinetic Measurements on Methylidyne Radical Reactions with Several Hydrocarbons at Low Temperatures. *Phys. Chem. Chem. Phys.* **2005,** *7* (15), 2921-2927.

41. Hickson, K. M.; Loison, J.-C.; Guo, H.; Suleimanov, Y. V. Ring-Polymer Molecular Dynamics for the Prediction of Low-Temperature Rates: An Investigation of the $C(^1D) + H_2$ Reaction. *J. Phys. Chem. Lett.* **2015,** *6*, 4194-4199.

42. Hickson, K. M.; Suleimanov, Y. V. An Experimental and Theoretical Investigation of the $C(^1D) + D_2$ Reaction. *Phys. Chem. Chem. Phys.* **2017,** *19* (1), 480-486.

43. Hickson, K. M.; Caubet, P.; Loison, J.-C. Unusual Low-Temperature Reactivity of Water: The $CH + H_2O$ Reaction as a Source of Interstellar Formaldehyde? *J. Phys. Chem. Lett.* **2013,** *4* (17), 2843-2846.

44. Vaghjiani, G. L. Kinetics of CH Radicals with $O_2$: Evidence for CO Chemiluminescence in the Gas Phase Reaction. *J. Chem. Phys.* **2003,** *119* (11), 5388-5396.

45. Hou, Z.; Bayes, K. D. Measurement of Rate Constants for Methylidyne(a$^4\Sigma$). *J. Phys. Chem.* **1992,** *96* (14), 5685-5687.

46. Hou, Z.; Bayes, K. D. Rate Constants for the Reaction of CH(a$^4\Sigma$) with NO, N$_2$, N$_2$O, CO, CO$_2$, and H$_2$O. *J. Phys. Chem.* **1993,** *97*, 1896-1900.

47. Bocherel, P.; Herbert, L. B.; Rowe, B. R.; Sims, I. R.; Smith, I. W. M.; Travers, D. Ultralow-temperature Kinetics of CH(X$^2\Pi$) Reactions: Rate Coefficients for Reactions with O$_2$ and NO (T=13-708 K), and with NH$_3$ (T=23-295 K). *J. Phys. Chem.* **1996,** *100* (8), 3063-3069.

48. Brownsword, R. A.; Canosa, A.; Rowe, B. R.; Sims, I. R.; Smith, I. W. M.; Stewart, D. W. A.; Symonds, A. C.; Travers, D. Kinetics over a Wide Range of Temperature (13-744 K): Rate Constants for the Reactions of CH($v$=0) with H$_2$ and D$_2$ and for the Removal of CH($v$=1) by H$_2$ and D$_2$. *J. Chem. Phys.* **1997,** *106* (18), 7662-7677.

49. Fleurat-Lessard, P.; Rayez, J.-C.; Bergeat, A.; Loison, J.-C. Reaction of Methylidyne Radical with CH$_4$ and H$_2$S: Overall Rate Constant and Absolute Atomic Hydrogen Production. *Chem. Phys.* **2002,** *279*, 87-99.

50. Butler, J. E.; Fleming, J. W.; Goss, L. P.; Lin, M. C. Kinetics of CH Radical Reactions with Selected Molecules at Room Temperature. *Chem. Phys.* **1981,** *56*, 355-365.

51. Bourgalais, J.; Capron, M.; Kailasanathan, R. K. A.; Osborn, D. L.; Hickson, K. M.; Loison, J.-C.; Wakelam, V.; Goulay, F.; Le Picard, S. D. The C($^3$P) + NH$_3$ Reaction in Interstellar Chemistry. I. Investigation of the Product Formation Channels. *Astrophys. J.* **2015,** *812* (2), 106.

52. Hickson, K. M.; Loison, J.-C.; Bourgalais, J.; Capron, M.; Le Picard, S. D.; Goulay, F.; Wakelam, V. The C($^3$P) + NH$_3$ Reaction in Interstellar Chemistry. II. Low Temperature Rate Constants and Modeling of NH, NH$_2$, and NH$_3$ Abundances in Dense Interstellar Clouds. *Astrophys. J.* **2015,** *812* (2), 107.

53. Shannon, R. J.; Cossou, C.; Loison, J.-C.; Caubet, P.; Balucani, N.; Seakins, P. W.; Wakelam, V.; Hickson, K. M. The Fast C($^3$P) + CH$_3$OH Reaction as an Efficient Loss Process for Gas-phase Interstellar Methanol. *RSC Adv.* **2014,** *4* (50), 26342.

54. Vastel, C.; Ceccarelli, C.; Lefloch, B.; Bachiller, R. Abundance of HOCO$^+$ and CO$_2$ in the Outer Layers of the L1544 Prestellar Core. *Astron. Astrophys.* **2016,** *591*, L2.

55. Dartois, E. The Ice Survey Opportunity of ISO. *Space Sci. Rev.* **2005,** *119* (1-4), 293-310.

56.     Ruaud, M.; Loison, J. C.; Hickson, K. M.; Gratier, P.; Hersant, F.; Wakelam, V. Modelling Complex Organic Molecules in Dense Regions: Eley-Rideal and Complex Induced Reaction. *Mon. Not. R. Astron. Soc.* **2015,** *447* (4), 4004-4017.

**Supporting information for**

**Tunneling Enhancement of the Gas-Phase CH + CO$_2$ Reaction at Low Temperature**

Dianailys Nuñez-Reyes,[1] Kevin M. Hickson,[1,*] Jean-Christophe Loison,[1] Rene F. K. Spada,[2] Rafael M. Vichietti,[3] Francisco B. C. Machado,[3] Roberto L. A. Haiduke[4]

[1]*Univ. Bordeaux, ISM, CNRS UMR 5255, F-33400 Talence, France.*
[2]*Departamento de Física, Instituto Tecnológico de Aeronáutica, 12228-900, São José dos Campos, SP, Brazil.*
[3]*Departamento de Química, Instituto Tecnológico de Aeronáutica, 12228-900, São José dos Campos, SP, Brazil.*
[4]*Departamento de Química e Física Molecular, Instituto de Química de São Carlos, Universidade de São Paulo, 13566-590, São Carlos, SP, Brazil.*

**This PDF files includes:**

Tables S1-S4 and Figure S1.

## Supplementary Tables

**Table S1** Forward rate constants (in cm$^3$particle$^{-1}$s$^{-1}$) for the CH + CO$_2$ → IM1 elementary step calculated at the ICVT and ICVT/SCT levels for different temperatures (in K) and by using different approaches to build the reactional path.

| T [K] | ICVT[a] | ICVT/SCT[a] | ICVT[b] | ICVT/SCT[b] | ICVT[c] | ICVT/SCT[c] |
|---|---|---|---|---|---|---|
| 25 | 1.57 × 10$^{-12}$ | 6.75 × 10$^{-12}$ | 9.54 × 10$^{-14}$ | 4.88 × 10$^{-12}$ | 2.69 × 10$^{-14}$ | 4.15 × 10$^{-12}$ |
| 50 | 2.12 × 10$^{-12}$ | 3.45 × 10$^{-12}$ | 5.22 × 10$^{-13}$ | 2.43 × 10$^{-12}$ | 2.77 × 10$^{-13}$ | 2.09 × 10$^{-12}$ |
| 75 | 2.08 × 10$^{-12}$ | 2.62 × 10$^{-12}$ | 8.18 × 10$^{-13}$ | 1.85 × 10$^{-12}$ | 5.37 × 10$^{-13}$ | 1.60 × 10$^{-12}$ |
| 100 | 2.02 × 10$^{-12}$ | 2.31 × 10$^{-12}$ | 1.00 × 10$^{-12}$ | 1.65 × 10$^{-12}$ | 7.30 × 10$^{-13}$ | 1.44 × 10$^{-12}$ |
| 125 | 1.99 × 10$^{-12}$ | 2.17 × 10$^{-12}$ | 1.13 × 10$^{-12}$ | 1.58 × 10$^{-12}$ | 8.80 × 10$^{-13}$ | 1.39 × 10$^{-12}$ |
| 150 | 1.98 × 10$^{-12}$ | 2.11 × 10$^{-12}$ | 1.24 × 10$^{-12}$ | 1.58 × 10$^{-12}$ | 1.01 × 10$^{-12}$ | 1.40 × 10$^{-12}$ |
| 175 | 2.00 × 10$^{-12}$ | 2.09 × 10$^{-12}$ | 1.34 × 10$^{-12}$ | 1.60 × 10$^{-12}$ | 1.12 × 10$^{-12}$ | 1.44 × 10$^{-12}$ |
| 200 | 2.04 × 10$^{-12}$ | 2.11 × 10$^{-12}$ | 1.44 × 10$^{-12}$ | 1.65 × 10$^{-12}$ | 1.23 × 10$^{-12}$ | 1.49 × 10$^{-12}$ |
| 225 | 2.10 × 10$^{-12}$ | 2.15 × 10$^{-12}$ | 1.54 × 10$^{-12}$ | 1.71 × 10$^{-12}$ | 1.33 × 10$^{-12}$ | 1.56 × 10$^{-12}$ |
| 250 | 2.16 × 10$^{-12}$ | 2.21 × 10$^{-12}$ | 1.63 × 10$^{-12}$ | 1.79 × 10$^{-12}$ | 1.44 × 10$^{-12}$ | 1.63 × 10$^{-12}$ |
| 275 | 2.24 × 10$^{-12}$ | 2.28 × 10$^{-12}$ | 1.73 × 10$^{-12}$ | 1.87 × 10$^{-12}$ | 1.55 × 10$^{-12}$ | 1.72 × 10$^{-12}$ |
| 300 | 2.32 × 10$^{-12}$ | 2.36 × 10$^{-12}$ | 1.84 × 10$^{-12}$ | 1.96 × 10$^{-12}$ | 1.66 × 10$^{-12}$ | 1.81 × 10$^{-12}$ |
| 325 | 2.42 × 10$^{-12}$ | 2.45 × 10$^{-12}$ | 1.95 × 10$^{-12}$ | 2.05 × 10$^{-12}$ | 1.77 × 10$^{-12}$ | 1.91 × 10$^{-12}$ |
| 350 | 2.52 × 10$^{-12}$ | 2.55 × 10$^{-12}$ | 2.06 × 10$^{-12}$ | 2.16 × 10$^{-12}$ | 1.88 × 10$^{-12}$ | 2.01 × 10$^{-12}$ |
| 375 | 2.61 × 10$^{-12}$ | 2.64 × 10$^{-12}$ | 2.17 × 10$^{-12}$ | 2.26 × 10$^{-12}$ | 1.99 × 10$^{-12}$ | 2.11 × 10$^{-12}$ |
| 400 | 2.71 × 10$^{-12}$ | 2.74 × 10$^{-12}$ | 2.28 × 10$^{-12}$ | 2.36 × 10$^{-12}$ | 2.10 × 10$^{-12}$ | 2.21 × 10$^{-12}$ |
| 500 | 3.18 × 10$^{-12}$ | 3.20 × 10$^{-12}$ | 2.77 × 10$^{-12}$ | 2.83 × 10$^{-12}$ | 2.60 × 10$^{-12}$ | 2.68 × 10$^{-12}$ |
| 700 | 4.38 × 10$^{-12}$ | 4.39 × 10$^{-12}$ | 3.96 × 10$^{-12}$ | 4.01 × 10$^{-12}$ | 3.79 × 10$^{-12}$ | 3.85 × 10$^{-12}$ |
| 1000 | 6.72 × 10$^{-12}$ | 6.73 × 10$^{-12}$ | 6.27 × 10$^{-12}$ | 6.30 × 10$^{-12}$ | 6.07 × 10$^{-12}$ | 6.12 × 10$^{-12}$ |
| 1500 | 1.20 × 10$^{-11}$ | 1.20 × 10$^{-11}$ | 1.14 × 10$^{-11}$ | 1.15 × 10$^{-11}$ | 1.12 × 10$^{-12}$ | 1.12 × 10$^{-12}$ |
| 2000 | 1.88 × 10$^{-11}$ | 1.88 × 10$^{-11}$ | 1.82 × 10$^{-11}$ | 1.82 × 10$^{-11}$ | 1.79 × 10$^{-12}$ | 1.79 × 10$^{-12}$ |
| 2500 | 2.73 × 10$^{-11}$ | 2.73 × 10$^{-11}$ | 2.65 × 10$^{-11}$ | 2.66 × 10$^{-11}$ | 2.62 × 10$^{-11}$ | 2.62 × 10$^{-11}$ |
| 3000 | 3.74 × 10$^{-11}$ | 3.74 × 10$^{-11}$ | 3.65 × 10$^{-11}$ | 3.65 × 10$^{-11}$ | 3.61 × 10$^{-11}$ | 3.62 × 10$^{-11}$ |
| 3500 | 4.90 × 10$^{-11}$ | 4.90 × 10$^{-11}$ | 4.81 × 10$^{-11}$ | 4.81 × 10$^{-11}$ | 4.76 × 10$^{-11}$ | 4.77 × 10$^{-11}$ |
| 4000 | 6.23 × 10$^{-11}$ | 6.23 × 10$^{-11}$ | 6.12 × 10$^{-11}$ | 6.12 × 10$^{-11}$ | 6.07 × 10$^{-11}$ | 6.07 × 10$^{-11}$ |

[a] The dual-level approach was employed using the electronic results of the CCSD(T)/CBS method.

[b] The dual-level approach was employed using the electronic results of the DKH-CCSD(T)/CBS method.

[c] The dual-level approach was employed using the electronic results of the DKH-CV-CCSD(T)/CBS method.

**Table S2** Forward rate constants (in cm$^3$particle$^{-1}$s$^{-1}$) for the CD + CO$_2$ → IM1 elementary step calculated at the ICVT and ICVT/SCT levels for different temperatures (in K) and by using different approaches to build the reactional path.

| T [K] | ICVT[a] | ICVT/SCT[a] | ICVT[b] | ICVT/SCT[b] | ICVT[c] | ICVT/SCT[c] |
|---|---|---|---|---|---|---|
| 25 | 2.53 × 10$^{-9}$ | 2.53 × 10$^{-9}$ | 3.37 × 10$^{-10}$ | 3.37 × 10$^{-10}$ | 9.49 × 10$^{-11}$ | 9.49 × 10$^{-11}$ |
| 50 | 9.77 × 10$^{-11}$ | 9.77 × 10$^{-11}$ | 2.41 × 10$^{-11}$ | 2.41 × 10$^{-11}$ | 1.28 × 10$^{-11}$ | 1.28 × 10$^{-11}$ |
| 75 | 2.33 × 10$^{-11}$ | 2.33 × 10$^{-11}$ | 9.17 × 10$^{-12}$ | 9.17 × 10$^{-12}$ | 6.01 × 10$^{-12}$ | 6.01 × 10$^{-12}$ |
| 100 | 1.13 × 10$^{-11}$ | 1.13 × 10$^{-11}$ | 5.60 × 10$^{-12}$ | 5.60 × 10$^{-12}$ | 4.08 × 10$^{-12}$ | 4.08 × 10$^{-12}$ |
| 125 | 7.34 × 10$^{-12}$ | 7.34 × 10$^{-12}$ | 4.19 × 10$^{-12}$ | 4.19 × 10$^{-12}$ | 3.25 × 10$^{-12}$ | 3.25 × 10$^{-12}$ |
| 150 | 5.58 × 10$^{-12}$ | 5.58 × 10$^{-12}$ | 3.50 × 10$^{-12}$ | 3.50 × 10$^{-12}$ | 2.83 × 10$^{-12}$ | 2.83 × 10$^{-12}$ |
| 175 | 4.66 × 10$^{-12}$ | 4.66 × 10$^{-12}$ | 3.12 × 10$^{-12}$ | 3.12 × 10$^{-12}$ | 2.61 × 10$^{-12}$ | 2.61 × 10$^{-12}$ |
| 200 | 4.13 × 10$^{-12}$ | 4.13 × 10$^{-12}$ | 2.91 × 10$^{-12}$ | 2.91 × 10$^{-12}$ | 2.48 × 10$^{-12}$ | 2.48 × 10$^{-12}$ |
| 225 | 3.81 × 10$^{-12}$ | 3.81 × 10$^{-12}$ | 2.79 × 10$^{-12}$ | 2.79 × 10$^{-12}$ | 2.42 × 10$^{-12}$ | 2.42 × 10$^{-12}$ |
| 250 | 3.61 × 10$^{-12}$ | 3.61 × 10$^{-12}$ | 2.73 × 10$^{-12}$ | 2.73 × 10$^{-12}$ | 2.40 × 10$^{-12}$ | 2.40 × 10$^{-12}$ |
| 275 | 3.50 × 10$^{-12}$ | 3.50 × 10$^{-12}$ | 2.71 × 10$^{-12}$ | 2.71 × 10$^{-12}$ | 2.42 × 10$^{-12}$ | 2.42 × 10$^{-12}$ |
| 300 | 3.44 × 10$^{-12}$ | 3.44 × 10$^{-12}$ | 2.73 × 10$^{-12}$ | 2.73 × 10$^{-12}$ | 2.45 × 10$^{-12}$ | 2.45 × 10$^{-12}$ |
| 325 | 3.43 × 10$^{-12}$ | 3.43 × 10$^{-12}$ | 2.76 × 10$^{-12}$ | 2.76 × 10$^{-12}$ | 2.51 × 10$^{-12}$ | 2.51 × 10$^{-12}$ |
| 350 | 3.44 × 10$^{-12}$ | 3.44 × 10$^{-12}$ | 2.82 × 10$^{-12}$ | 2.82 × 10$^{-12}$ | 2.58 × 10$^{-12}$ | 2.58 × 10$^{-12}$ |
| 375 | 3.48 × 10$^{-12}$ | 3.48 × 10$^{-12}$ | 2.89 × 10$^{-12}$ | 2.89 × 10$^{-12}$ | 2.66 × 10$^{-12}$ | 2.66 × 10$^{-12}$ |
| 400 | 3.54 × 10$^{-12}$ | 3.54 × 10$^{-12}$ | 2.97 × 10$^{-12}$ | 2.97 × 10$^{-12}$ | 2.75 × 10$^{-12}$ | 2.75 × 10$^{-12}$ |
| 500 | 3.92 × 10$^{-12}$ | 3.92 × 10$^{-12}$ | 3.41 × 10$^{-12}$ | 3.41 × 10$^{-12}$ | 3.20 × 10$^{-12}$ | 3.20 × 10$^{-12}$ |
| 700 | 5.09 × 10$^{-12}$ | 5.09 × 10$^{-12}$ | 4.61 × 10$^{-12}$ | 4.61 × 10$^{-12}$ | 4.40 × 10$^{-12}$ | 4.40 × 10$^{-12}$ |
| 1000 | 7.55 × 10$^{-12}$ | 7.55 × 10$^{-12}$ | 7.04 × 10$^{-12}$ | 7.04 × 10$^{-12}$ | 6.82 × 10$^{-12}$ | 6.82 × 10$^{-12}$ |
| 1500 | 1.32 × 10$^{-11}$ | 1.32 × 10$^{-11}$ | 1.26 × 10$^{-11}$ | 1.26 × 10$^{-11}$ | 1.23 × 10$^{-11}$ | 1.23 × 10$^{-11}$ |
| 2000 | 2.06 × 10$^{-11}$ | 2.06 × 10$^{-11}$ | 1.99 × 10$^{-11}$ | 1.99 × 10$^{-11}$ | 1.96 × 10$^{-11}$ | 1.96 × 10$^{-11}$ |
| 2500 | 2.98 × 10$^{-11}$ | 2.98 × 10$^{-11}$ | 2.89 × 10$^{-11}$ | 2.89 × 10$^{-11}$ | 2.86 × 10$^{-11}$ | 2.86 × 10$^{-11}$ |
| 3000 | 4.07 × 10$^{-11}$ | 4.07 × 10$^{-11}$ | 3.97 × 10$^{-11}$ | 3.97 × 10$^{-11}$ | 3.93 × 10$^{-11}$ | 3.93 × 10$^{-11}$ |
| 3500 | 5.33 × 10$^{-11}$ | 5.33 × 10$^{-11}$ | 5.23 × 10$^{-11}$ | 5.23 × 10$^{-11}$ | 5.18 × 10$^{-11}$ | 5.18 × 10$^{-11}$ |
| 4000 | 6.77 × 10$^{-11}$ | 6.77 × 10$^{-11}$ | 6.65 × 10$^{-11}$ | 6.65 × 10$^{-11}$ | 6.60 × 10$^{-11}$ | 6.60 × 10$^{-11}$ |

[a] The dual-level approach was employed using the electronic results of the CCSD(T)/CBS method.

[b] The dual-level approach was employed using the electronic results of the DKH-CCSD(T)/CBS method.

[c] The dual-level approach was employed using the electronic results of the DKH-CV-CCSD(T)/CBS method.

**Table S3** Thermochemical properties (in kJ/mol) for the CH(CD) + $CO_2$ → IM1 elementary step.[a]

| | CH + $CO_2$ → IM1 | | | |
|---|---|---|---|---|
| **Method** | **ΔE** | **ΔH** | **V$^‡$** | **ΔV$_a^{G,‡}$** |
| CCSD(T)/CBS[b] | -171.1 | -153.8 | -6.4 | 0.4 |
| DKH-CCSD(T)/CBS[c] | -168.5 | -151.3 | -5.8 | 1.1 |
| DKH-CV-CCSD(T)/CBS[d] | -170.8 | -153.5 | -5.6 | 1.3 |
| | CD + $CO_2$ → IM1 | | | |
| **Method** | **ΔE** | **ΔH** | **V$^‡$** | **ΔV$_a^{G,‡}$** |
| CCSD(T)/CBS[b] | -171.1 | -156.9 | -6.4 | -1.4 |
| DKH-CCSD(T)/CBS[c] | -168.5 | -154.4 | -5.8 | -0.8 |
| DKH-CV-CCSD(T)/CBS[d] | -170.8 | -156.6 | -5.6 | -0.5 |

[a] The equilibrium geometries considered for these calculations were previously calculated with the UCCSD/cc-pVDZ method.[10] The properties presented are the classical barrier (V$^‡$, electronic energy difference between the reactants and the saddle point), the adiabatic barrier height (ΔV$_a^{G,‡}$, defined as V$^‡$ + ΔZPE$^‡$), the electronic energy of the reaction (ΔE), the enthalpy of the reaction at 0 K (ΔH = ΔE + ΔZPE).
[b] The CBS extrapolation was calculated by using cc-pVXZ (X=4 and 5) basis sets.
[c] The CBS extrapolation was calculated by using cc-pVXZ-DK (X=4 and 5) basis sets.
[d] The CBS extrapolation was calculated by using cc-pCVXZ-DK (X=4 and 5) basis sets.

**Table S4** Continuous supersonic flow characteristics

| Mach number | 1.8 ± 0.02[a] | 2.0 ± 0.03 | 3.0 ± 0.1 | 3.9 ± 0.1 |
|---|---|---|---|---|
| Carrier gas | $N_2$ | Ar | Ar | Ar |
| Density (× $10^{16}$ cm$^{-3}$) | 9.4 | 12.6 | 14.7 | 25.9 |
| Impact pressure (Pa) | 1093.2 | 1399.9 | 2039.8 | 3946.3 |
| Stagnation pressure (Pa) | 1373.2 | 1853.2 | 4653.0 | 15065.4 |
| Temperature (K) | 177 ± 2[a] | 127 ± 2 | 75 ± 2 | 50 ± 1 |
| Mean flow velocity (ms$^{-1}$) | 496 ± 4[a] | 419 ± 3 | 479 ± 3 | 505 ± 1 |

[a]The errors on the Mach number, temperature and mean flow velocity, cited at the level of one standard deviation from the mean are calculated from separate measurements of the impact pressure using a Pitot tube as a function of distance from the Laval nozzle and the stagnation pressure within the reservoir.

**Supplementary Figures**

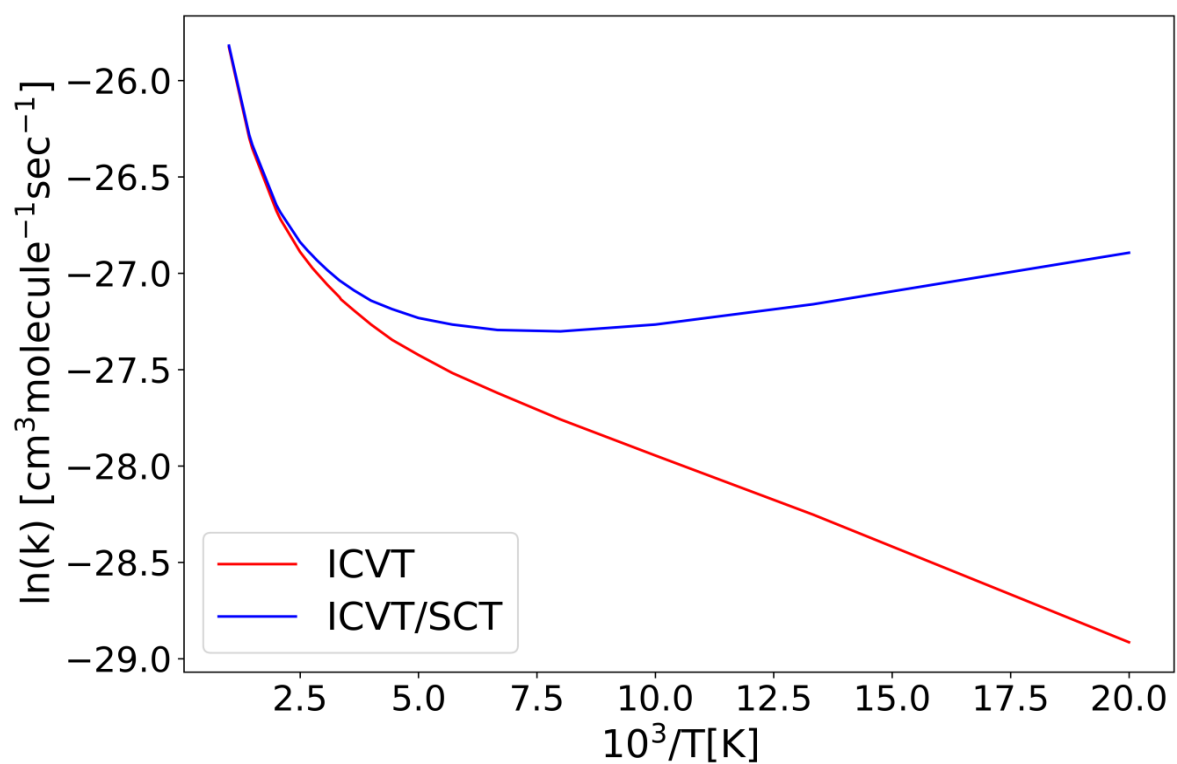

**Figure S1.** Quantum tunneling effects (ICVT/SCT versus ICVT) on calculated rate constants of the CH + $CO_2$ reaction (see Table S1).